\documentclass{aa}
\usepackage{natbib}
\usepackage{xspace}
\usepackage{amssymb}
\usepackage{amsmath}
\usepackage{graphicx}
\usepackage{comment}

\def\aap{A\& A}

\def\mnras{MNRAS}
\def\nat{Nature}
\def\apj{ApJ}

\def\apjl{ApJL}
\def\apjs{ApJS}

\def\St{\ensuremath{\mathrm{St}}\xspace}

\newcommand{\se}[1]{\mbox{Sect.\ \ref{sec:#1}}}

\newcommand{\eq}[1]{\mbox{Eq.\ (\ref{eq:#1})}}

\newcommand{\fg}[1]{\mbox{Fig.\ \ref{fig:#1}}}
\newcommand{\Fg}[1]{\mbox{Figure\ \ref{fig:#1}}}
\newcommand{\Tb}[1]{\mbox{Table\ \ref{tab:#1}}}
\newcommand{\tb}[1]{\mbox{Tab.\ \ref{tab:#1}}}

\newcommand{\eg}{e.g.,}

\newcommand{\mr}[1]{\mathrm{#1}}
\newcommand{\ff}{\phi}
\newcommand{\tff}{$\phi$ }
\newcommand{\ep}[1]{\cdot 10^{#1}}

\newcommand{\figlarge}[3]{
\begin{figure*}[!ht]
\centering
\includegraphics[width=15cm]{#1}
\caption{#2}
\label{fig:#3}
\end{figure*}
}

\newcommand{\figsmall}[3]{
\begin{figure}[!ht]
\centering
\includegraphics[width=9cm]{#1}
\caption{#2}
\label{fig:#3}
\end{figure}
}

\begin{document}
%\thesaurus{02.01.2,08.03.4,08.06.2,08.16.5,13.09.6}
%\title{Can residual infall replenish the small particle population in a disk?}
\title{The structure of dust aggregates in hierarchical coagulation}
\titlerunning{Hierarchical aggregates} \authorrunning{Dominik, Paszun,
  Borel} \author{C.~Dominik$^{1}$, D. Paszun, \& H. Borel
  \thanks{This paper was originally submitted to A\&A in June 2009.
    Despite a positive report by an anonymous referee, it was never
    published since the referee asked for an analytical derivation of
    some of the results.  D. Paszun had left astronomy by then, so we
    never did comply with this request, even though it should be
    possible to do that derivation.  I (C. Dominik) am putting it up
    on arXiv now because the results have renewed relevance in
    relation to Rosetta results \citep{2016Natur.537...73B}, and the
    various studies on pebble collapse as a model for comet and
    planetesimal formation
    \citep[e.g.][]{2016A&A...587A.128L,2014Icar..235..156B}.  This
    study was part of the PhD thesis of Dr. D. Paszun at the
    University of Amsterdam.  An electronic version of that PhD thesis
    is available at
    \texttt{http://www.astro.uva.nl/static/research/\newline{}\hspace*{5cm}theses/phd/dm-paszun.pdf}}}
\institute{$^{1}$Anton Pannekoek Institute for Astronomy, Science Park
  904, NL-1098 XH Amsterdam, The Netherlands; E--mail: dominik@uva.nl}
\date{DRAFT, \today}

\abstract{Dust coagulation in interstellar space and protoplanetary
  disks is usually treated as one of 2 extreme cases:
  \emph{Particle-Cluster Aggregation} and \emph{Cluster-Cluster
    Aggregation}.  In this paper we study the process of
  \emph{hierarchical growth}, where aggregates are built from
  significantly smaller aggregates (but \emph{not} monomers).  We show
  that this process can be understood as a modified, PCA-like process
  that produces porous, but non-fractal particles whose filling factor
  is chiefly determined by the porosity of the building blocks.
  We also show that in a coagulation environment where relative
  velocities are driven by turbulence, a logarithmically flat mass
  distribution (equal mass per mass decade) as it is typically found
  in environments where fragmentation replenishes small grains, leads
  to a situation where small particles and aggregates dominate the
  growth of large ones.  Therefore, in such environments, hierarchical
  growth should be seen as the norm.  Consequently, we predict that
  the aggregates in such environments are not fractals with extremely
  low densities as they would result from extrapolation fractal laws
  to large sizes.  The compactification of aggregates does not only
  result from collisions with enough energy to restructure aggregates
  - it starts already earlier by filling voids in particles with
  smaller grains that contribute to the growth.}

\maketitle


\section{\label{sec:ch2-introduction}Introduction}

The formation of planets proceeds in disks surrounding young stars.
Small dust grains collide and stick to each other growing to larger
and larger sizes.  This growth initially involves energies that are
insufficient to break or even deform aggregates (also referred to as
particles or agglomerates) made of smaller constituents - monomers
(also referred to as grains).  In the commonly accepted scenario, dust
grows to relatively large, at least dm-sized particles.  At these
sizes aggregates suffer from a very strong head wind (due to
sub-Keplerian motion of the gas) that leads to strong radial drift
\citep{1977MNRAS.180...57W}) and destructive collisions
\citep[e.g.][]{2008A&A...480..859B}.  These processes are critical to
our understading of planet formation, and they depend strongly on the
porosity of the dust grains \citep{2007A&A...469.1169B}. 

The growth of aggregates from small monomers is usually treated in one
of two limiting cases:  Particle-Cluster Aggregation (PCA) and
Cluster-Cluster Aggregation (CCA).  These terms are sometimes used
loosely.  PCA means the formation of an aggregate by a succession of
collisions with individual monomers.  Each collision can be seen as an
independent collision, the impactors always come in as individual
particles and do not collide with other impactors on the way in.  CCA
means the build-up of a large aggregate in a sequence of collisions of
aggregates of similar sizes.  In its purest form, the two colliding
aggregates will always have exactly the same size, in reality they
will be similar in size.

It has been shown that the structure of the aggregates formed by PCA
and by CCA, respectively, is very different.

PCA aggregates get a homogeneous structure.  The core of the aggregate
has a constant density and is therefore not fractal in nature.  In
full generality this has been shown by \citet{1984PhRvA..29.2966B} who
demonstrated that the PCA buildup is subject to a causality condition
that requires a lower limit for the fractal dimension $D_\mr{f}$ of
the forming aggregate given by $D_\mr{f}\ge
D_\mr{space}-D_\mr{approach}+1$ where $D_\mr{space}$ is the dimension
of space and $D_\mr{approach}$ is the fractal dimension of the
trajectory of the approaching particle.  For ballistic approach
trajectories, we have $D_\mr{approaching}=1$, and in normal 3D space,
this means that $D_\mr{f}\ge 3$.  Therefore, pure PCA aggregates are
non-fractal.

The case of CCA aggregates has been studied in the laboratory
\citep[e.g.][]{2004PhRvL..93b1103K} and theoretically
\citep{1999Icar..141..388K,2006Icar..182..274P} as well, and it has
been shown that such growth leads to much smaller fractal dimensions.
The details depend again on the structure of the collision trajectory
of the particles and on the rotational state of the colliding
aggregates.

Physically, PCA and CCA aggregation represent very different growth
conditions.  PCA corresponds to a runaway growth situation, where one
or a small number of particles consume all the other particles before
they have time to aggregate themselves.  This means that the typical
timescale for a particle to collide with the runaway aggregate must be
shorter than the time to collide with any monomer in the environment.
At least in a situation where the aggregation process starts with only
small particles present, this is a very unrealistic situation.
Therefore, in practice, at least initially CCA is the much more likely
situation.  A good example is the growth of aggregates under
conditions where relative motions are governed by thermal motions, e.g
Brownian motion of particles in a rarefied gas.  Such growth has
recently been studied under microgravity conditions and indeed it was
shown that this process leads to a narrow size distribution
\citep{2004PhRvL..93b1103K}, implying that the typical collision takes
place between two aggregates of similar size.  Consequently and as
expected, such experiments show the buildup of fractal aggregates,
with fractal dimensions around 1.5.  As long as the collision
velocities are largest between the smallest grains, the growth process
will deplete the smallest particles by collisions with other small
particles and in this way keep the size distribution narrow.  CCA-like
growth is the result and will continue until the basic physics of the
growth process change.  The fractal structure of the aggregates is not
even a requirement for growth to proceed in a CCA-like way.  As long
as small particles are removed efficiently, the size distribution will
remain compact and collisions of similar-sized particles will be
dominant in determining the growth process.

There are two basic conditions under which the growth process will
leave the CCA path and we will call these conditions the
\emph{runaway} and the \emph{fragmentation} case.

In the runaway case, the conditions change to those also needed for
PCA growth: the relative motions of particles become such that indeed
the most likely collision for a small particle or aggregate is the
collision with a large particle.  In protoplanetary disks, such
conditions may arise when systematic motions start to govern the
relative motions rather than thermal motions.  For example, under the
conditions of turbulent motions in the gas, we can have two kinds of
particles. Small particles that are very well coupled to the gas
follow a similar motion and thus their relative velocities remain
small
\citep{1980A&A....85..316V,1993prpl.conf.1031W,2007A&A...466..413O}.
The dust-gas interaction is quantified by the Stokes number $St =
\tau_\mr{f}/\tau_\mr{L}$, where $\tau_\mr{f}$ is the stopping or
friction time of a particle and $\tau_\mr{L}$ is the overturn time of
the largest eddies (usually assumed to be of the order of the Kepler
time $\sim \Omega^{-1}$).  The stopping time of a particle is given by
\begin{equation}
\tau_\mr{f} = \frac{3}{4 c_\mr{s} \rho_\mr{g}} \frac{m}{\sigma}, 
\label{eq:tau-f}
\end{equation} 
where $c_\mr{s}$ is the sound speed in the gas, $\rho_\mr{g}$ is density of gas,
$m$ is the mass of a dust particle, and $\sigma$ is the projected surface of the
particle.  Larger particles with Stokes numbers close to unity will start to
decouple from the gas and find themselves in a sand-blasting stream of gas
containing small, well-coupled particles.

The second case is that of \emph{fragmentation}.  It was shown by
\citet{2005A&A...434..971D,2008A&A...491..663D} that the observed
abundance of small grains in protoplanetary disks is actually best
explained by ongoing fragmentation that destroys aggregates of a
certain size and puts the mass of these back into small grains.  In
such a case, small grains will always contribute to the growth and in
this way introduce a hierarchical growth component.

In order to understand the evolution of porosity during the
coagulation of dust particles, it is therefore important to understand
the effects of hierarchical growth.  In this paper, we will study
hierarchical growth and describe the structure of aggregates formed in
this way.

\section{\label{sec:ch2-methods}Methods}

To study the structure of aggregates resulting from hierarchical
growth, we use the model developed by \citet{2006Icar..182..274P}.  In
this approach, aggregates are treated as rigid bodies that freely move
and rotate in 3-dimensional space.  Aggregates can move simply
ballistically on trajectories that lead to collisions, or they may be
embedded in a medium causing both motion and rotation to change
frequently (after one stopping length) in order to model Brownian
motion.  In this study we will concentrate on the low density limit in
which the mean Brownian path length of an aggregate is larger that the
size of the aggregate, so that the motion can be seen as ballistic
during the collision.  \citet{2006Icar..182..274P} had shown that
under high density conditions in the innermost parts of the solar
nebula this condition may not hold (\eg\ aggregates made of 10$^3$
micron-sized monomers at densities above $\rho_\mr{g} > 10^{-9}$
g/cm$^3$ have a mean free path shorter than their own size), but it is
valid throughout a large part of the solar nebula including the
formation region of the Earth.  A basic assumption of this model is
that aggregates do not restructure, i.e. that any contact made between
grains stays forever and cannot be moved.  Physically, this is
equivalent to the assumption that collision energies are much lower
than the energy required to initiate restructuring.  This energy is
given by the rolling energy
\begin{equation}
E_\mr{roll}=6 \pi^2 \gamma R \xi_\mr{crit}.
\label{eq:Erol}
\end{equation}
In \eq{Erol}, $\gamma$ is the surface energy, $R$ is the reduced
radius of two monomers in contact, and $\xi_\mr{crit}$ is the critical
displacement needed to initiate irreversible rolling. We refer the
reader to \citet{1997ApJ...480..647D,1993ApJ...407..806C} for detailed
description of the contact physics.

In \citet{2006Icar..182..274P} this model has been used to collide
aggregates of equal size and is capable to simulate small aggregates,
which was sufficient for the comparison with zero-gravity laboratory
experiments \citep{2004PhRvL..93b1103K}.  To study the effects of
hierarchical growth, aggregates must be built that consist of millions
of monomers.  To speed up the search for new contacts between
colliding aggregates, we have implemented an efficient nearest
neighbor search (NNS) algorithm to improve the performance of the
model \citep{1988csup.book.....H}.  The algorithm we are using is
optimized for the special setup of the model: there is no need to
search for nearest neighbors within each aggregate, but only between
grains of different aggregates.  Therefore, we keep the particle lists
for each aggregate separate. With the nearest neighbor search, the
dependence of the computation time on the number of monomers is now
reduced from ${\cal O}(N_1\times N_2)$ to ${\cal O}(N_2 \times
\log(N_1))$.

The computations are very efficient and allow the construction of
aggregates with up to 10$^7$ monomers.  The computational bottleneck
is no longer the collision search, but the rotation of the large
target aggregate which requires $N$ matrix operations to compute
positions of the monomers for each time step.  However, it turns out
that we can safely ignore rotation for the present study, because the
rotation of the large aggregate will quickly become negligible.  In
order to demonstrate this, lets assume that the kinetic energy
$E$ governing the random motions of the aggregates (for example
through Brownian motion) remains constant during the growth.  As the
mass of the projectiles remains constant during the hierarchical
growth process, we keep the approach velocity of a small projectile
$v_\mr{p}$ constant as well.  As the target becomes larger, its own
random motions can be ignored and the approach velocity is dominated
by the velocity of the small projectile.  In order to see if the
rotation of the target is important, we need to compare the linear
approach velocity with the velocity of the outer regions of the target
caused by its rotation.  The angular velocity of a target particle is
given as
\begin{equation}
\omega_\mr{t} = \sqrt{\frac{2 E}{I}}=\sqrt{\frac{2 E}{C m_0 N R^2}},
\label{eq:ang-vel}
\end{equation}
where $I$ is the moment of inertia, $m_0$ is a monomer mass, and the constant
$C$ depends on a geometry of an aggregate (e.g. for a sphere $C=2/5$).  The
equivalent circumferential speed is then
\begin{equation}
v_\mr{cir,t} \propto \sqrt{\frac{1}{N}}
\label{eq:circ-vel-N}
\end{equation}
and decreases with increasing mass.  Even if the approach velocity is
comparable to the velocity caused by rotation initially, when target
and projectile are of similar size, the ratio of the two velocities
quickly approaches zero as the target is growing.  Therefore, we will
conduct most of this study ignoring rotation during approach.  We
will, however, confirm the validity of this assumption with a test
calculation (see \se{ch2-rot}).  Within a few collision timescales the
rotation of the target aggregate effectively stops and only rotation
of projectiles can be considered.  We will discuss this effect
separately in \se{ch2-discussion}.

A schematic picture of a collision configuration is shown in
\figsmall{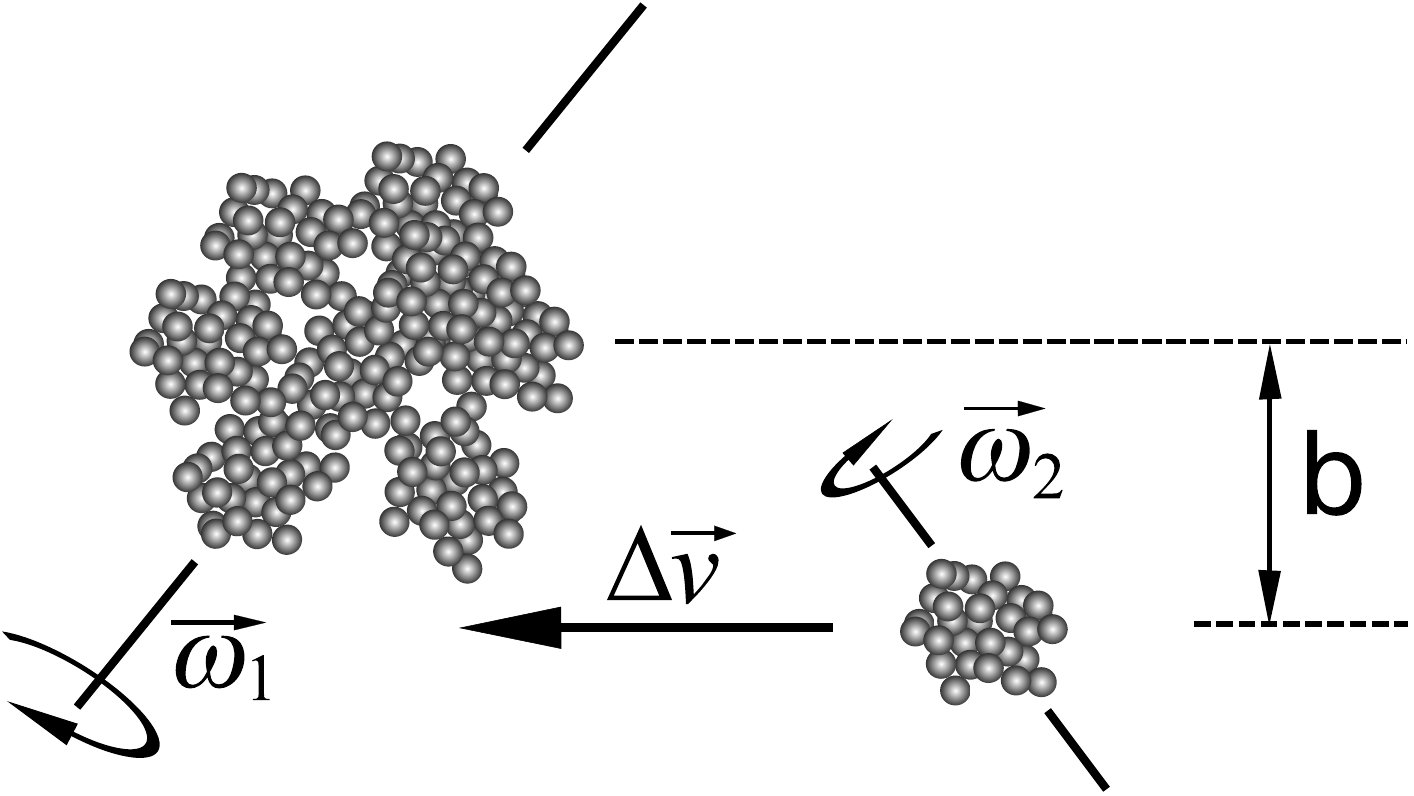}{A sketch of a
  collision. Aggregates rotate around a random axis and approach at
  random impact parameter $b$. The relative velocity $\Delta v$ is
  assumed to be in the hit-and-stick regime.}{sample} \fg{sample}.  A
target aggregate initially is of the same size as the incoming
projectile.  As the growth progresses, the size of the target
increases.  Thus the process starts as a perfect case of the CCA
aggregation and smoothly turns into the PCA-like hierarchical case.
In the cases where we do treat rotation, both particles spin around
random axes with rotation velocities $\omega_i$.  All velocities are
drawn from the Maxwell distribution.  The temperature is assumed to be
300 K and monomers are 1 micron (in diameter) sized silica spheres.

To quantify the structure of aggregates produced by hierarchical
growth, we use the filling factor within a sphere enclosing the entire
aggregate
\begin{equation}
\ff = N \Bigl(\frac{r_0}{r_\mr{out}}\Bigr)^3,
\label{eq:ff}
\end{equation}
where $N$ is number of particles in an aggregate, $r_0$ is the monomer
radius, and $r_\mr{out}$ is the outer radius of the aggregate.

PCA particles generally have a uniform density structure, meaning that
the filling factor is constant throughout the core of an aggregate.
The density will not be constant throughout the entire aggregate
because of an inevitable transition region in the outer parts of the
aggregate - we will discuss this region in more detail in
\se{ch2-surf-effect}.  CCA aggregates, on the other hand, are known to
have a fractal structure resulting in a density decreasing from the
core to the outer regions with a power law dependence
\begin{equation}
\rho(r) \propto r^{D_\mr{f}-3},
\label{eq:rho_fractal}
\end{equation}
where $D_\mr{f}$ is the fractal dimension and is about 1.5 for the
ideal CCA process \citep{2004PhRvL..93b1103K,2006Icar..182..274P}.

To study the effects of projectile size, we grow aggregates by
sequential addition of particles of constant mass.  We sample over two
orders of magnitude in projectile mass. Our smallest projectiles
are monomers, and the largest are built of 256 grains.  The intermediate
masses are successive powers of two in monomer mass
($2^0m_0,2^2m_0,\ldots,2^8m_0$).

It is to be expected that the internal structure of the projectiles will also
have influence on the resulting aggregates.  We cover this parameter by
considering two extreme cases for growing the projectiles.  In the first case,
projectiles are made by PCA aggregation, leading to projectiles with an upper
limit (reached only in the case of large aggregates) for the filling factor of
$\ff=0.15$ \citep{1992A&A...263..423K}.  The real aggregates used for this
simulations have lower filling factors, typically $\ff \approx 0.05$.

%
%
% RESULTS
%
%
\section{\label{sec:ch2-results}Results}

We present the internal structure of 16 aggregates produced in the
hierarchical growth using projectiles of different structure and
mass. \Tb{agg-mass} shows the final size of the grown aggregates in
units of the projectile mass.  Note that in terms of number of
monomers, the largest aggregates are the ones made of large
projectiles, reaching about $6.5\times10^{6}$ monomers for the
aggregates grown from 256-mers.  However, we give the mass in units of
the projectile mass because the degree to which the structural limit
of hierarchical growth can be reached is determined by this number
rather than the number of monomers.  The growth of largest particles
is limited to 25000 projectiles.
\begin{table}[!ht]
\caption{Mass of aggregates presented in this paper normalized to projectile
  mass. $^a$ - Simulation of hierarchical growth with rotation taken into
  account.}  \label{tab:agg-mass}
\begin{tabular}{c r r r}
\hline\hline
$N_\mr{p}$ & \multicolumn{3}{c}{final $N/N_\mr{p}$} \\
 & \multicolumn{3}{c}{aggregates made of bullet type} \\
& \multicolumn{1}{c}{PCA} & \multicolumn{1}{c}{CCA} & \multicolumn{1}{c}{CCA$^a$} \\
\hline
$2^0$ &  $497\ 500$  &   $497\ 500$ & $40\ 300$ \\
$2^1$ &  $328\ 500$  &   $328\ 500$ & $11\ 400$ \\
$2^2$ &  $296\ 250$  &   $198\ 900$ & $6\ 080$  \\
$2^3$ &  $210\ 000$  &   $125\ 000$ & $3\ 840$  \\
$2^4$ &  $140\ 625$  &   $112\ 150$ & $1\ 880$  \\
$2^5$ &  $87\ 500$   &   $60\ 290$  & $1\ 090$  \\
$2^6$ &  $81\ 500$   &   $42\ 000$  & $750$     \\
$2^7$ &  $45\ 625$   &   --         & $460$     \\
$2^8$ &  $25\ 900$   &   $26\ 500$  & $245$     \\
\hline
\end{tabular}
\end{table}

%
%
% RESULTS: PCA
%
%
\subsection{\label{sec:ch2-pca}Non-fractal (PCA) projectiles}

The simplest case of the hierarchical growth is the sequential
addition of monomers onto a larger target (the pure PCA case).  This
process has been studied before and we present it as a reference case
in this study.  Aggregates produced from PCA projectiles of different
mass are presented in \fg{PCA-img-nr}.  The reference particle is
shown in \fg{PCA-img-nr}a.  Its
\figlarge{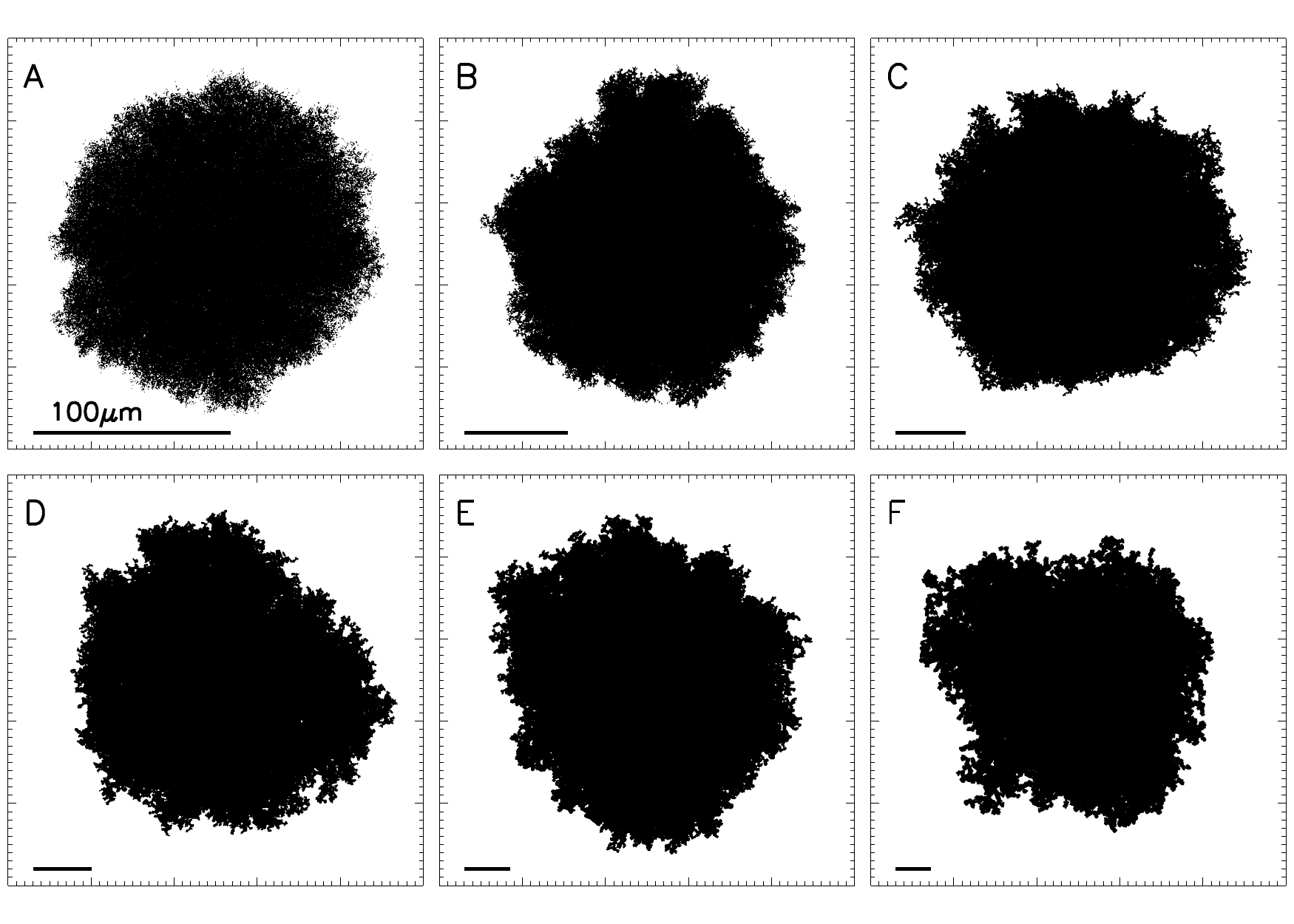}{Aggregates produced by a sequential
  coagulation of small PCA projectiles of a constant size. Each panel presents a
  horizontal bar indicating length of 100 $\mu$m. Projectile mass is:
  A-$N_\mr{p}=2^0$, B-$N_\mr{p}=2^2$, C-$N_\mr{p}=2^4$, D-$N_\mr{p}=2^5$,
  E-$N_\mr{p}=2^6$, F-$N_\mr{p}=2^8$.}{PCA-img-nr}
physical size is indicated by a scale bar that represents the length of 100
$\mu$m, i.e. 100 monomer diameters.  This particle consists of approximately $N
\approx 5\ep{5}$ monomers (cf. \tb{agg-mass}).  The remaining aggregates shown in
the figure are grown from projectiles of increasing mass.  They contain more
monomers, but the number of impactors forming these aggregates decreases with
increasing projectile mass.  The largest aggregate consists of over $N=6 \ep{6}$
monomers (\fg{PCA-img-nr}f), but merely $\sim 26000$ projectiles (each consisting
of 256 grains).

\figsmall{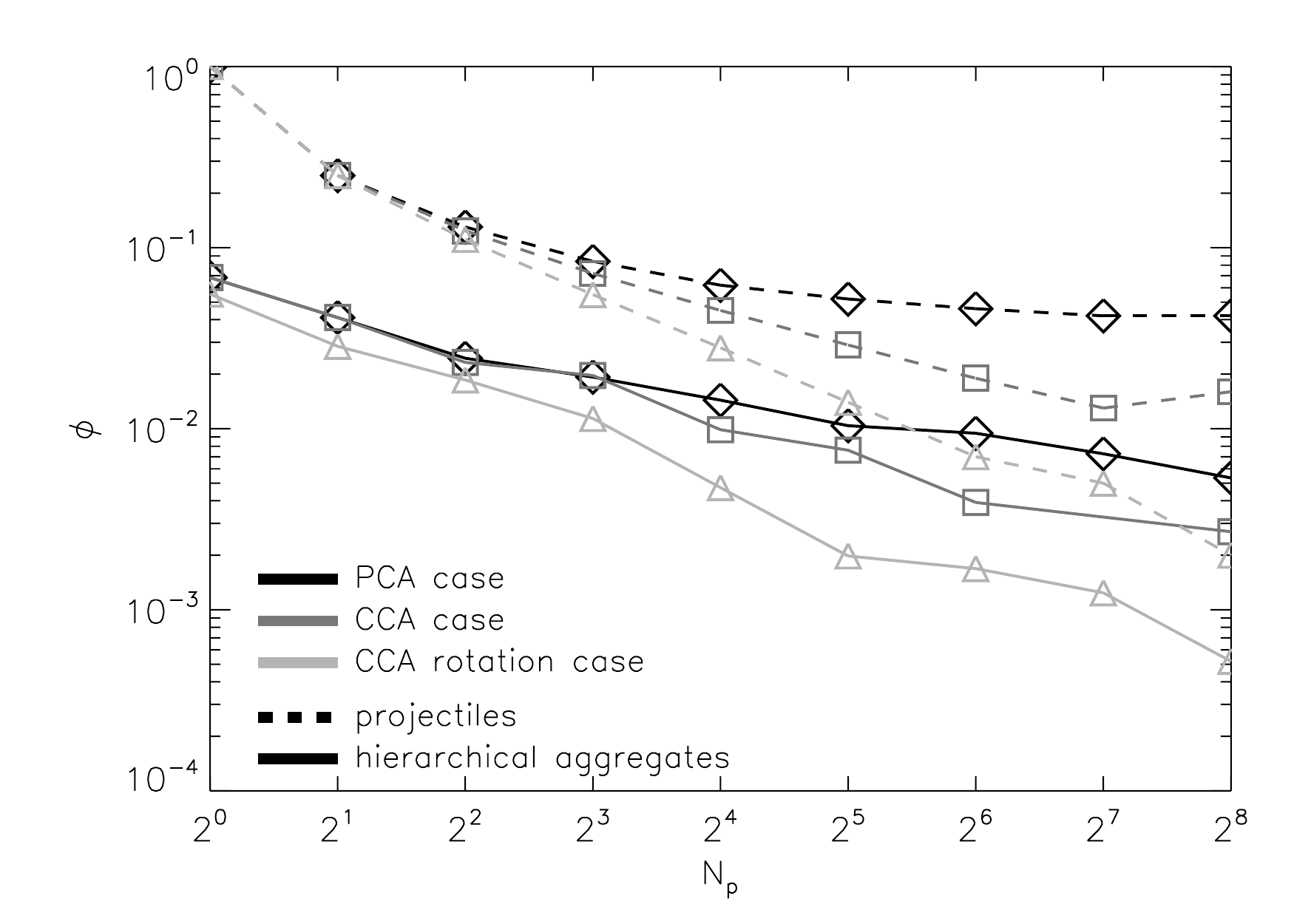}{The filling factor of aggregates
  formed by the hierarchical growth as a function of projectile size. The solid
  lines represent filling factor of final aggregates, while the dashed lines
  show the filling factor of projectiles.Black lines correspond to PCA case,
  while the grey lines to the CCA case.}{ff-vs-np}

The mass difference between the largest (\fg{PCA-img-nr}f) and the
smallest (\fg{PCA-img-nr}a) aggregates is about one order of
magnitude. The physical size also differs about one order of
magnitude, which already indicates that the internal structure of
agglomerates produced by the hierarchical growth depends on projectile
size.  This can directly be seen by inspecting the outer layers of all
particles, where the size of voids increases with increasing
projectile size (cf. \fg{PCA-img-nr}a -- \fg{PCA-img-nr}f).  Also the
overall filling factor decreases as the projectile size
increases. This behavior is shown quantitatively in \fg{ff-vs-np},
along with the filling factors of the projectiles themselves.  The
filling factor of the PCA projectiles decreases from one (for
monomers) to about 0.04 (for 256-mers) and clearly stabilizes at about
this value.  The filling factor of the aggregates formed by
hierarchical growth (solid black line) decreases by over an order of
magnitude as the PCA projectile mass increases from 1 to 256 monomers.
Moreover, the decrease of the overall filling factor does not slow
down as in the case of projectiles. This decrease for large,
hierarchically grown aggregates is caused by two effects.  The main
effect is due to the decreasing filling factor of the projectiles.
Since they can be considered as porous grains, their coagulation
directly leads to a lower filling factor than in the pure monomer PCA
case.  However, as the filling factor of the projectiles is stabilizing
near 0.04, one might expect that also the filling factor of the
aggregates grown from these particles should stabilize, a result that
is not evident in \fg{ff-vs-np}.  As we will see below, this is caused
by the comparatively large transition region close to the surface of
the aggregate.  For the largest projectiles, the computation has not
advanced far enough to make the aggregate core dominate the overall
filling factor.  For a better view on the internal structure, we show
in \fg{density-shells-nr} the packing density $\ff (r)$ as a function
of distance from the center of mass of an aggregate. The filling
factor is roughly constant
\figsmall{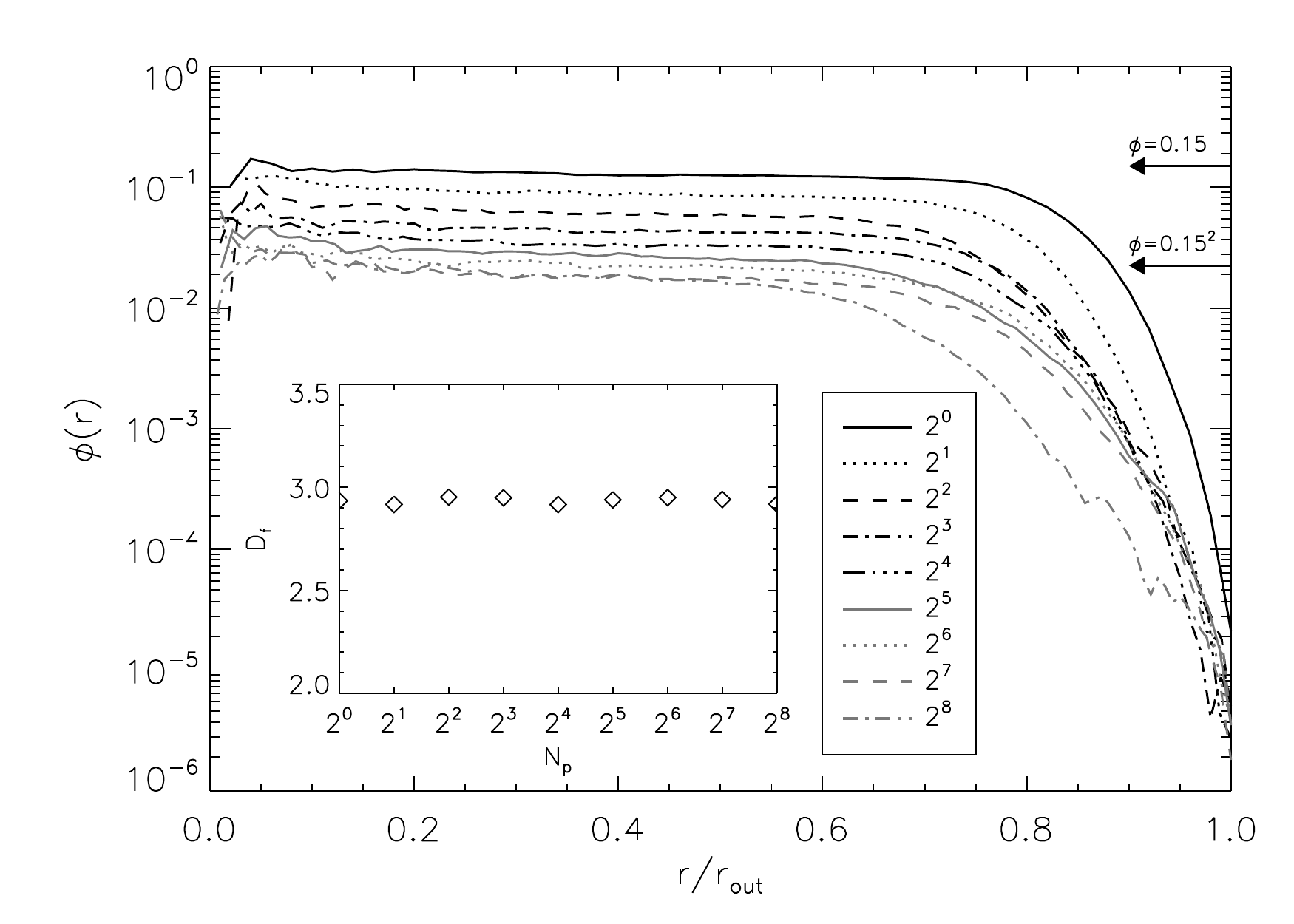}{Density structure of aggregates
  formed by a sequential incorporation of small PCA bullets. Different lines
  correspond to aggregates formed out of projectiles of different masses. From
  top to bottom, projectile masses are: $N_\mr{p}=2^0,2^1,2^2,\ldots,2^8$. The
  inset shows the fractal dimension $D_\mr{f}$ determined for these aggregates
  as a function of the mass of a projectile.}{density-shells-nr}
throughout the inner region of aggregates. Small fluctuations indicate
local inhomogeneities due to the finite size of monomers.  These local
density variations increase with increasing projectile mass, an effect
of the larger building units and their own internal structure.  The
core density structure of two aggregates made of the largest
projectiles ($N_\mr{p}=128$ and $N_\mr{p}=256$) is very similar.  In
combination with the result that the projectile filling factors were
very similar, this indicates that the resulting density of an
aggregate formed by hierarchical growth is mainly a function of the
filling factor of the building components.  In fact, filling factor in
the inner region has the value of about $\ff \approx 0.02$, not far
from the value $\ff = 0.0225 = 0.15^2$.  This latter value would be
expected from an idealized two-step aggregation process.  The
formation of the projectiles from monomers through PCA growth causes a
filling factor of 0.15.  If we assume that these aggregates are
spherical as well, the following build-up of the target from these
projectile aggregates lowers the filling factor by the same factor
again.  What is interesting here is that the filling factors of our
projectiles were actually lower (0.04), due to the fluffy transition
area at the projectile surface.  From the simple argument above, we
would expect a filling factor of $\ff=0.04*0.15=0.006$.  The fact that
the true value is much closer to $0.15^2$ shows that, apparently, the
second growth step does compensate the low projectile filling factor
by a certain degree of geometrical penetration. Since projectiles have
fluffy surface regions, the first contact occurs after the projectile
has penetrated a bit into the surface region of the target, filling
some of the voids created by the projectile formation.  This result is
further discussed in \se{ch2-tooth}.

The uniform density throughout the core of aggregates indicates that their
structure is non-fractal. To verify this, we determine the fractal dimension
$D_\mr{f}$ of our aggregates by fitting the data with a power-law function
\begin{equation}
N(r_\mr{g}) = K \biggl(\frac{r_\mr{g}}{r_0}\biggr)^{D_\mr{f}}, 
\label{eq:Df}
\end{equation}
where $K$ is the fractal pre-factor and $r_\mr{g}$ is the radius of
gyration defined as
\begin{equation}
r_\mr{g} = \sqrt{\frac{\sum_i^N r_i^2}{N}},
\label{eq:r-gyr}
\end{equation}
with $r_i$ being a distance of $i$-th grain from the center of mass.
This gyration radius is calculated for an aggregate at different
stages of its growth sequence.  Since the structure of the projectiles
influences the fit at small $N(r_\mr{g})$, the power-law exponent
$D_\mr{f}$ is fitted to the outer half of the mass of an aggregate.
The resulting fractal dimensions of different particles are shown in
an inset of \fg{density-shells-nr}.  The fractal dimension oscillates
between $D_\mr{f}=2.9$ and $D_\mr{f}=3.0$, which shows the non-fractal
structure of particles, as expected for the PCA case
\citep{1984PhRvA..29.2966B}.

%
%
% RESULTS: CCA
%
%
\subsection{\label{sec:ch2-cca}Fractal (CCA) projectiles}
CCA growth produces porous, fractal aggregates
\citep{2004PhRvL..93b1103K,2006Icar..182..274P}. The filling factor of these
particles depends on mass (due to fractal structure) and is lower than that of
PCA clusters of equal mass.  This is a direct result of the CCA process, in
which the void volume is increased with every growth step.  The hierarchical
growth with these fluffy projectiles is expected to produce more porous
particles than in the case of PCA bullets.

The Results of hierarchical growth by a sequential agglomeration of
CCA particles are presented in \fg{CCA-img-nr}.
\figlarge{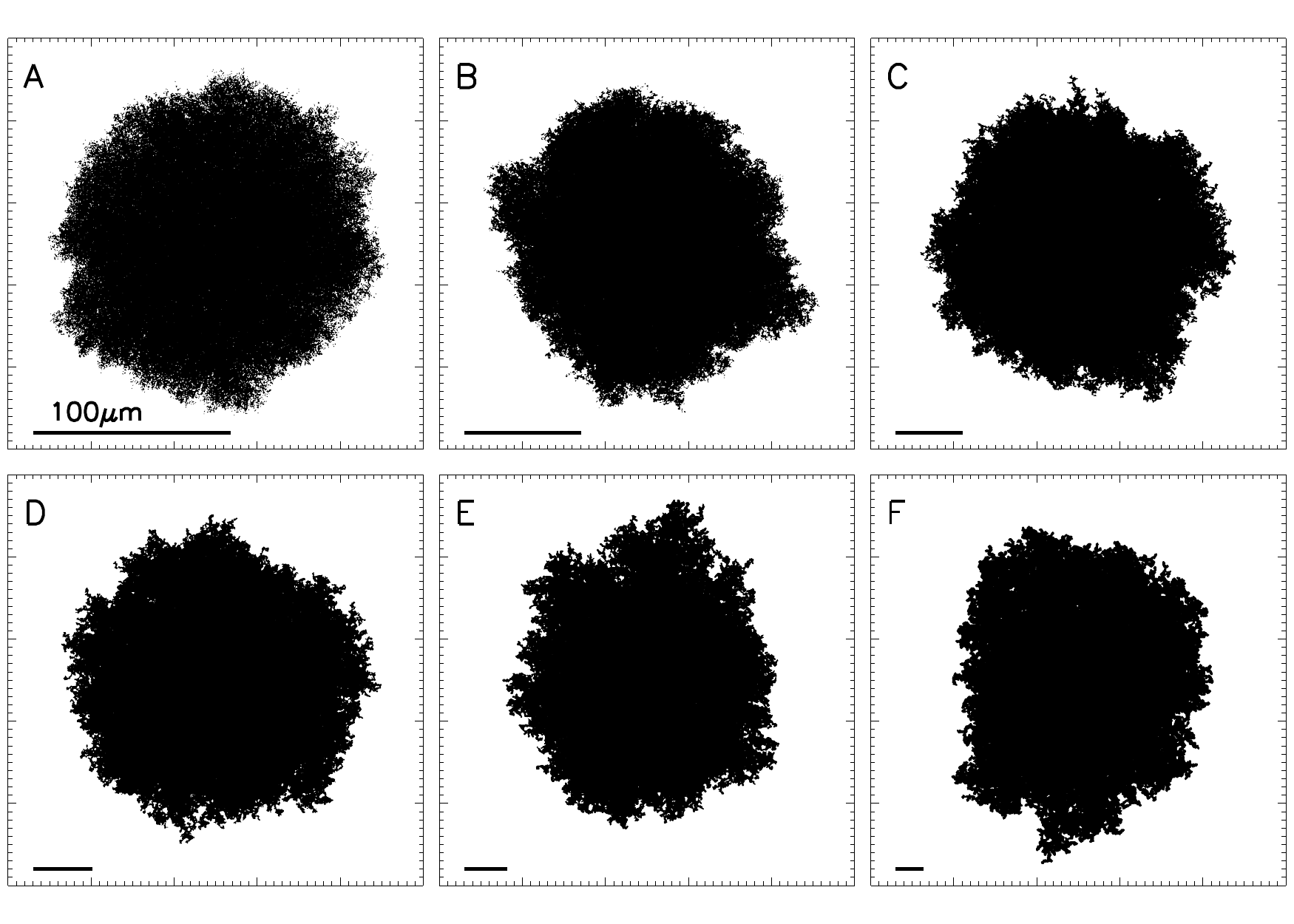}{Aggregates produced by a sequential
  coagulation of small CCA projectiles of a constant size. Each panel presents a
  horizontal bar indicating a length of 100 $\mu$m. Projectile mass is:
  A-$N_\mr{p}=2^0$, B-$N_\mr{p}=2^2$, C-$N_\mr{p}=2^4$, D-$N_\mr{p}=2^5$,
  E-$N_\mr{p}=2^6$, F-$N_\mr{p}=2^8$.}{CCA-img-nr}
At first sight, in comparison with the PCA bullet case, these
aggregates show no spectacular differences.  All particles have an
approximately spherical shape and indicate an increase of porosity
with increasing bullet mass. In \fg{CCA-img-nr}a we present the
\emph{reference} aggregate made by a sequential accumulation of
monomers. Note that this is just a copy of the aggregate in
\fg{PCA-img-nr}a.  Even though the total mass of aggregates made of
fractal (CCA) bullets is lower (cf. \tb{agg-mass}), the size (as
indicated by the bar in each panel of \fg{CCA-img-nr}) is very similar
to or even larger than in the case of PCA projectiles. Obviously, the
global filling factor of the new aggregates is much lower
now. Quantitatively this is shown in \Fg{ff-vs-np}.  The solid, grey
line indicates the filling factor of aggregates produced by the
hierarchical growth with CCA projectiles.  For monomer and dimer
\emph{bullets}, the agglomerates are obviously identical since the
bullets are identical as well.  However, even for quadrumer and
octomer bullets, the filling factors of the final aggregates are
identical both for CCA and PCA bullets.  However, this is a direct
consequence of the fact that the filling factors of bullets themselves
hardly differ from each other at these small sizes.  Starting at
$N_\mr{p} \ge 2^4$, the differences between the targets do become
significant and again simply reflect the changes in the bullet
properties.

Again, the filling factors plotted in \Fg{ff-vs-np} are influenced by
the fluffy transition region close to the aggregate surface.  A better
view at the limiting filling factors that are reached by this process
and realized in the cores of the aggregates is given in
\Fg{CCAdensity-shells-nr} which shows the density structure of
aggregates made of different CCA projectiles.
\figsmall{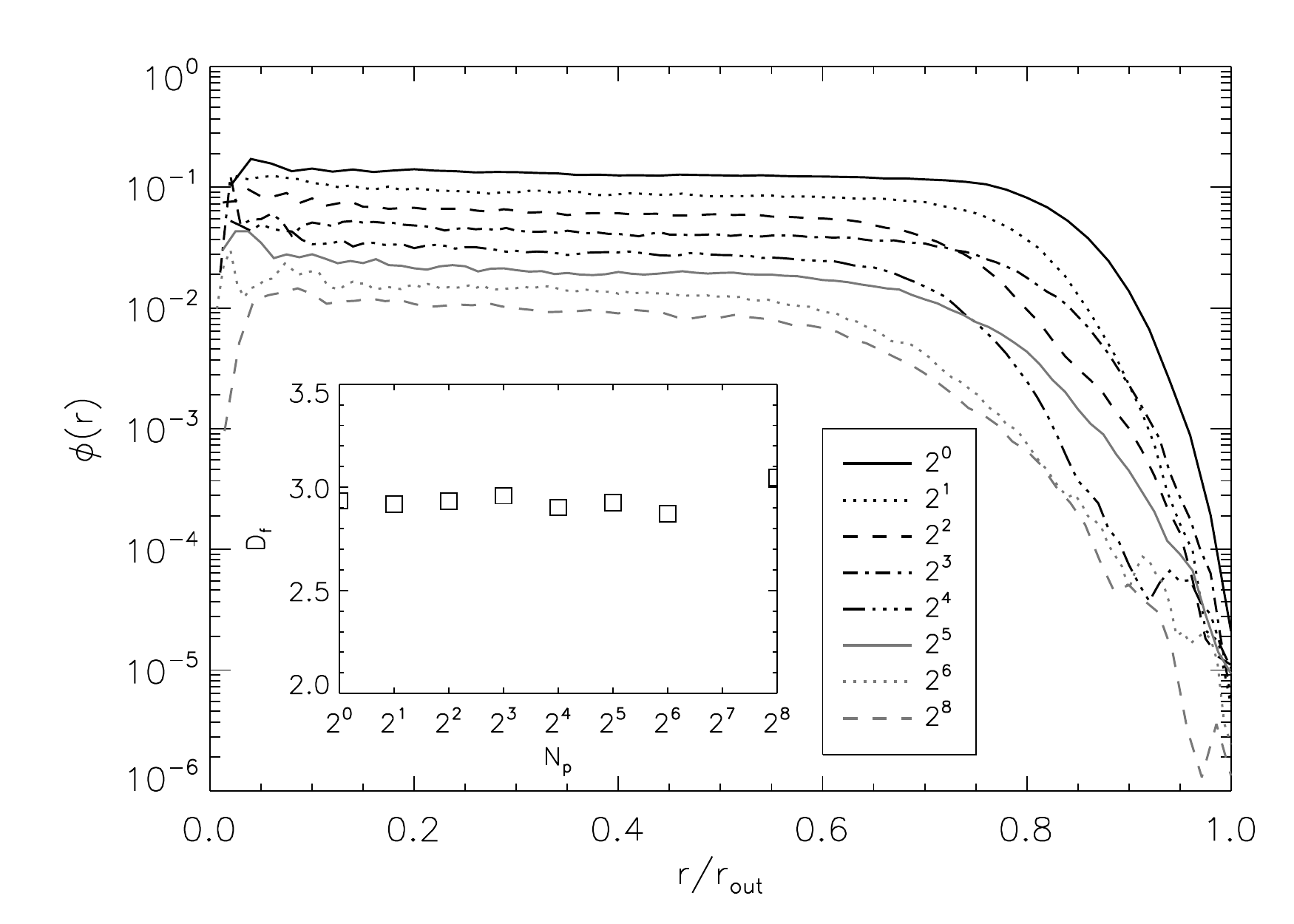}{Density structure of CCA aggregates formed
  without rotation. Different lines correspond to aggregates formed out of
  projectiles of different masses. From top to bottom, projectile masses are:
  $N_\mr{p}=2^0,2^1,2^2,\ldots,2^8$. The inset shows the fractal dimension
  $D_\mr{f}$ determined for these aggregates as a function of the mass of a
  projectile.}{CCAdensity-shells-nr} 
An increasing projectile size results in a decrease of the central
filling factor. For large CCA projectiles ($N_\mr{p}=2^5\ldots 2^8$)
the filling factor decreases below the values obtained in the case of
PCA projectiles (cf. \fg{density-shells-nr}).  The aggregate made of
the largest \emph{bullets} has a central filling factor $\ff=10^{-2}$,
about a factor 2 lower than the value obtained for the corresponding
PCA projectiles, due to the lower filling factor of the CCA
\emph{bullets}.  The more irregular structure of the CCA projectiles
also causes the stronger ``noise'' in the filling factor in the inner
regions of aggregates at $r<0.2r_\mr{out}$.  Even though the CCA
projectiles are fractals, this property is lost in larger aggregates
during the hierarchical growth. The filling factor in a final
agglomerate is approximately constant throughout the inner half of the
aggregate's radius and declines only in the outer, porous layers.
\emph{Therefore the internal structure of hierarchical aggregates made
  of fractal projectiles is homogeneous and non-fractal.} The inset in
\fg{CCAdensity-shells-nr} shows the fractal dimension of these
aggregates.  Clearly the initial fractal nature of projectiles
diminishes, as the fractal dimension determined for these hierarchical
aggregates is very close to $D_\mr{f}=3.0$.

%
%
% DISCUSSION:
%
%
\section{\label{sec:ch2-discussion}Discussion}

We have shown that hierarchical growth in the pure form which we have
studied here produces aggregates whose structure depends mostly on the
filling factor of the projectiles.  However several other effects may
also play a role.  Here we address the influence of these additional
factors.  The influence of the fractal surface region is addressed in
\se{ch2-surf-effect}.  The importance of aggregate penetration
(toothing) on the filling factor of the resulting aggregates is
discussed in \se{ch2-tooth}.  The effect of rotation is presented in
\se{ch2-rot} and illustrated with additional simulations.  Finally the
relevance of the hierarchical growth for coagulation of dust in
protoplanetary disks is discussed in \se{ch2-relevance}.

%
%
% DISCUSSION: SURFACE POROSITY
%
%
\subsection{\label{sec:ch2-surf-effect}Surface porosity}

PCA aggregates show a dependence of the global filling factor on the
aggregate mass.  The reason for this is that the filling factor is
affected by the fluffy, outer layers.  This \emph{transition region}
initially covers the entire particle. As growth proceeds, projectiles
are filling voids and build the constant-density core of an aggregate
with the filling factor $\ff = 0.15 \ff_\mr{p}$.  Eventually, the
density in the center of an aggregate is high enough to prevent
projectiles from entering. In the mean time, however, the aggregate
has grown and a new transition region has appeared near the surface.

We determine the influence of these outer layers on the global filling
factor as a function of aggregate mass.  To approach the limit of
$\ff=0.15$ for the pure PCA growth, the effect of the transition
region must decrease with an increasing aggregate
size. \Fg{transition} shows the density structure of a single PCA
aggregate
\figsmall{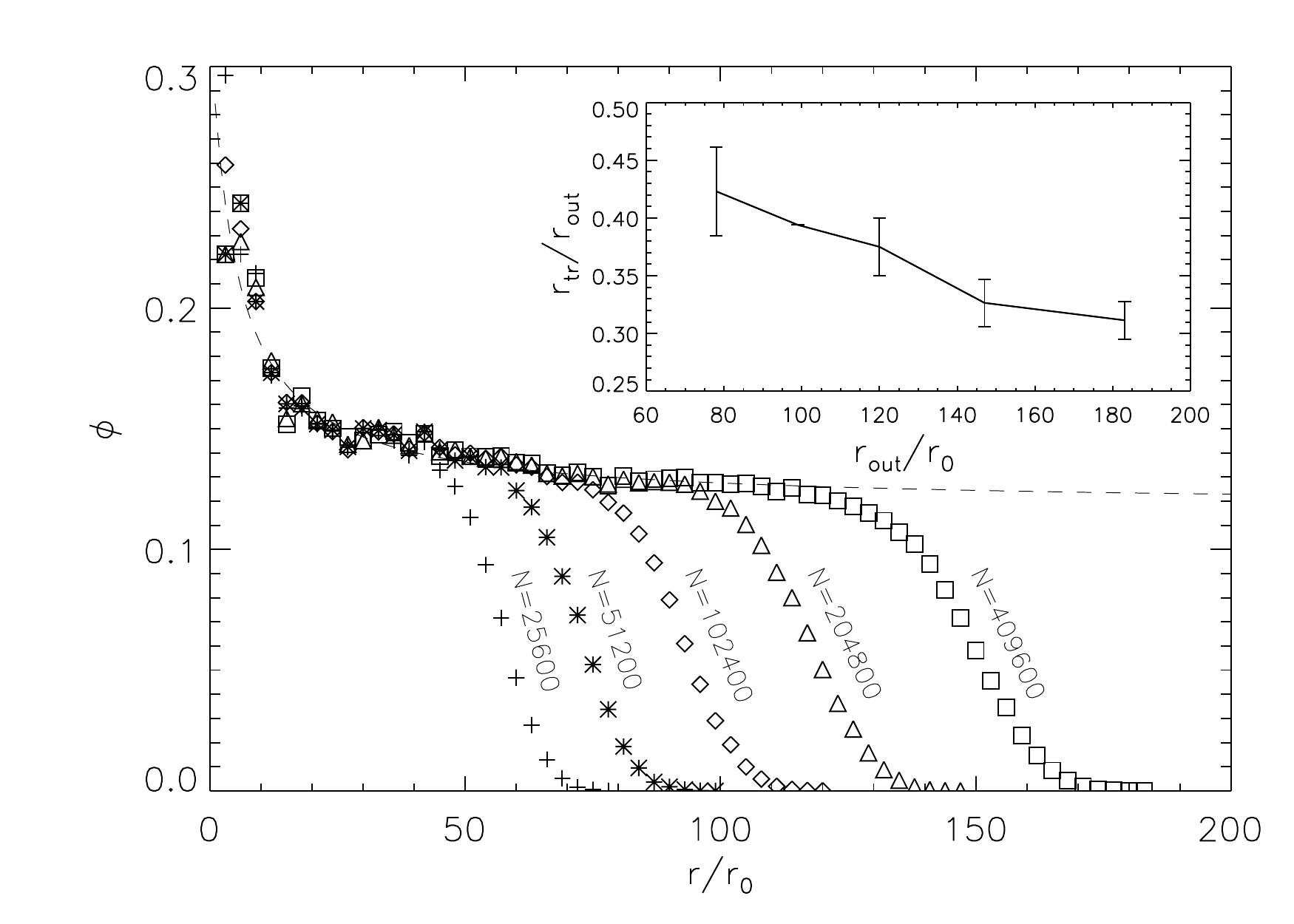}{The filling factor as a function of
  aggregate size. The growth sequence is shown at 5 different stages:
  $N=25600$, $N=51200$, $N=102400$, $N=204800$, $N=409600$ monomers. The dashed
  line shows a fit to the data of the inner part of the aggregate at the most
  advanced growth stage. The transition layer is defined here as the region,
  where the filling factor is lower than the fit by more than 10\%. The inset
  shows the relative size of the transition layer as a function of aggregate
  size. Again 5 different growth stages are indicated.}{transition}
obtained at different growth stages.  To determine the thickness of
the transition layer $r_\mr{tr}$, we fitted an arbitrary function
$f(r)=1/r+\phi_0$, with the fitting parameter $\phi_0$, to the inner
(core) region of an aggregate at different stages.  As the beginning of
the transition layer we assume a radius of an aggregate, where the
filling factor drops below the fitted function by $\Delta \phi=0.01$
($\sim10\%$).

Accumulation of monomers causes an increase of the outer radius
$r_\mr{out}$ but also causes an increase of the thickness of the outer
porous layer $r_\mr{tr}$. However, the increase of the transition
layer is slower than the increase of the aggregate size. This behavior
is shown in the inset of \fg{transition}. Indicated error-bars are
estimated by changing the criterion for the onset of the transition
regime by $10\%$. The thickness of the transition region relative to
the outer radius of an aggregate decreases as the growth
proceeds. Moreover we can show that the volume of the transition
region is decreasing. This transition region corresponds to a volume
of
\begin{equation}
V_\mr{tr} = \frac{4}{3} \pi (r_\mr{out}^3 - r_\mr{tr}^3),
\label{eq:transition-V}
\end{equation}
and when we divide \eq{transition-V} by the total volume $V=4/3 \pi
r_\mr{out}^3$, we find
\begin{equation}
\frac{V_\mr{tr}}{V} = 3\frac{r_\mr{tr}}{r_\mr{out}} -3 \biggl(
\frac{r_\mr{tr}}{r_\mr{out}} \biggr)^2 + \biggl( \frac{r_\mr{tr}}{r_\mr{out}}
\biggr)^3.
\label{eq:transition-Vrat}
\end{equation}
Thus the volume ratio decreases with a decreasing ratio of
$r_\mr{tr}/r_\mr{out}$.  Therefore the influence of the outer, fluffy
layers weakens with increasing size. In the limiting case, this outer,
fluffy region is relatively very thin leading to a filling factor of
about $\ff=0.15$.

%
%
% DISCUSSION: TOOTHING RADIUS
%
%
\subsection{\label{sec:ch2-tooth}The core porosity and the toothing radius}

The core density of hierarchical aggregates shows that their filling
factor results from PCA-like growth of porous particles.  Therefore,
the packing density is chiefly the product of the PCA filling factor
($\ff=0.15$) and the filling factor of a projectile. In the case of
PCA bullets the filling factor reaches the value of $\ff\approx 0.02$
(cf. \fg{density-shells-nr}), indicating penetration of projectiles
into the target (see \se{ch2-pca}).  The lower the filling factor of
the projectile, the larger the surface voids in the particles are, and
the larger the overlap is expected to be.
\figsmall{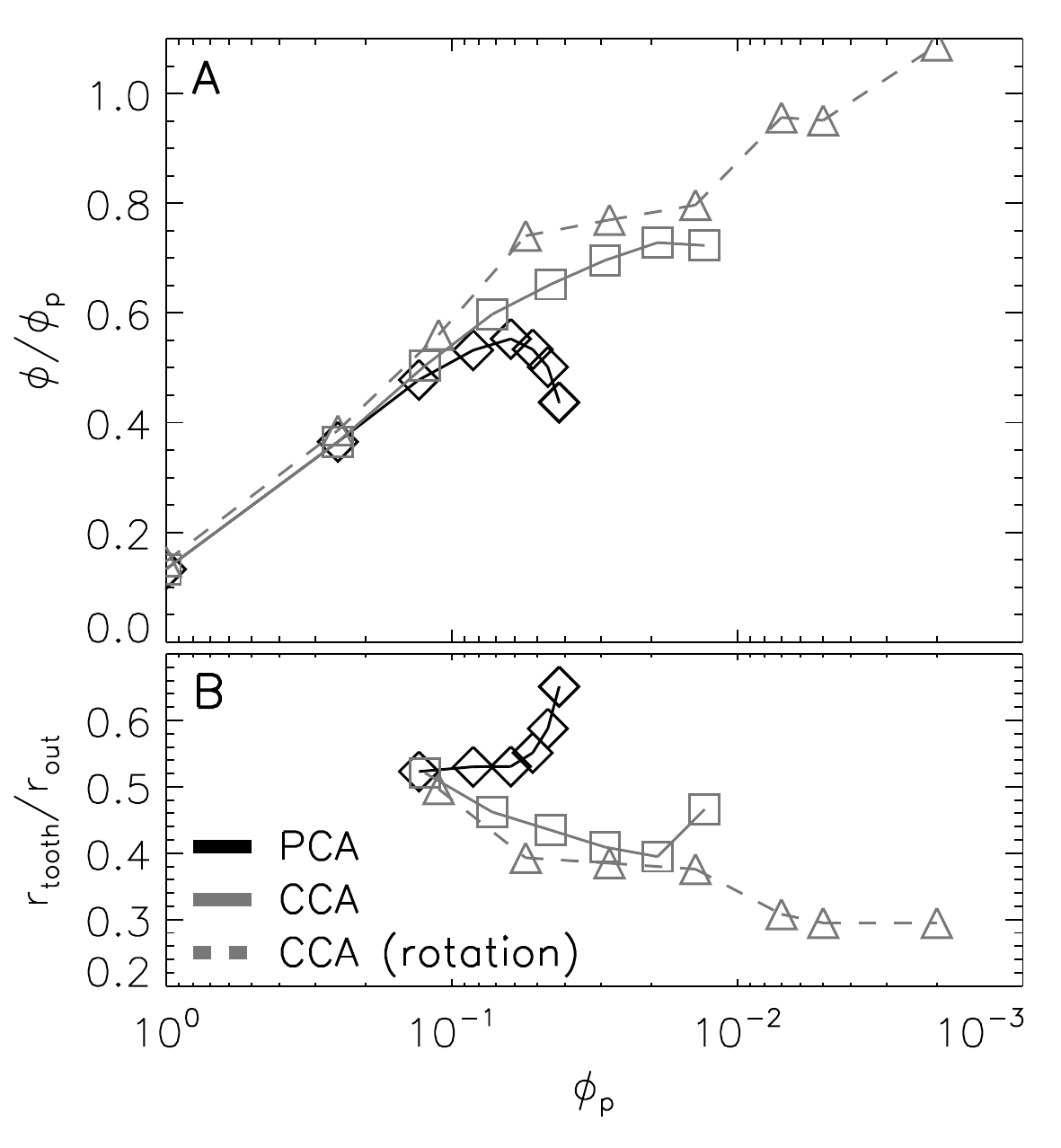}{A - The core filling factor
  of hierarchically grown aggregates normalized to the filling factor of
  projectiles. The core filling factor is determined by taking the filling
  factor of the inner part of an aggregate only. In this case we determined the
  packing density within a sphere with a radius of half of the outer radius. The
  dotted line indicate a relation fitted to the data. B- a toothing radius
  $r_\mr{tooth}$ of projectiles. See text for further explanation.}{tooth}
\Fg{tooth}a shows the ratio of aggregate over projectile filling
factor for aggregates formed from different projectiles. The deviation
from the $\ff/\ff_\mr{p}=0.15$ line is evidently increasing with
decreasing filling factor. The exception of PCA particles results from
the non-fractal structure of these projectiles. Their low filling
factor \tff is caused by the extended, outer transition
region. Therefore, a further increase in size of projectiles results
in an increase of the filling factor of PCA projectiles (because the
relative thickness of the transition region decreases, as shown in
\se{ch2-surf-effect}) and eventually reaches the limit of
$\ff/\ff_\mr{p}=0.15$ at $\ff_\mr{p}=0.15$. CCA projectiles on the
other hand have a fractal nature and thus their filling factor
decreases with increasing size.  This results in a continuous increase
of the ratio of the aggregate to the projectile filling factor, as the
intersection is increasingly larger. Therefore the filling factor of a
final aggregate can be presented as the PCA growth, where the final
filling factor is given by $\ff = 0.15 \ff_\mr{p}$ and a correction
due to the penetration and the resulting increase of the actual
density. The correction must be a function of the projectile porosity,
as more fluffy aggregates can be penetrate deeper. Following this path
we fitted a simple function to our data
\begin{equation}
\ff = 0.15\ \ff_\mr{p} - \zeta\ \ff_\mr{p} \log \ff_\mr{p},
\label{eq:phi-fit}
\end{equation}
where $\zeta = 0.368 \pm 0.012$ is a constant obtained from the least
square fit.

The penetration depth can be expressed in terms of a toothing radius
$r_\mr{tooth}$. This quantity has been introduced by
\citet{1993A&A...280..617O} as ``half of the distance of the centers
of two equal size clusters that are on average sticking to each
other''. Thus in our case the toothing radius $r_\mr{tooth}$ is given
by
\begin{equation}
\frac{\phi}{\phi_\mr{p}(r_\mr{tooth})}=0.15,
\end{equation}
where 
\begin{equation}
\phi_\mr{p}(r_\mr{tooth}) = N_\mr{p}(r_\mr{tooth}) \biggl(
\frac{r_0}{r_\mr{tooth}} \biggr)^3
\end{equation}
is the filling factor of a projectile determined at
$r=r_\mr{tooth}$. Therefore a low toothing radius means that the
penetration is deep and aggregates overlap significantly. \Fg{tooth}b
shows the toothing radii obtained for projectiles with different
filling factors. PCA particles are characterized by rather large
toothing radii, meaning that the penetration must be relatively
shallow. Intuitively, the toothing radius of non-fractal aggregates
should approach the outer radius in the limit of large agglomerates
(i.e. for PCA projectiles $\lim_{n\rightarrow \infty}
\frac{r_\mr{tooth}}{r_\mr{out}} = 1$). Here PCA particles still suffer
from the extended, porous, outer transition region. Thus the average
penetration depth is relatively deep ($r_\mr{tooth} / r_\mr{out} <
0.65$ for $N_\mr{p} \ge 2^7$). \Fg{tooth}b shows, however, that the
toothing radius increases very steeply for PCA particles. It may thus
be expected that for larger sizes (and filling factor \tff=0.15) the
toothing radius will approach the outer radius.

Fractal projectiles on the other hand show very different behavior. In
this case the penetration depth increases with the increasing size up
to 70\% of their outer radius for largest aggregates ($N_\mr{p}=2^8$).
Aggregates of different fractal dimension (CCA particles formed in the
presence of rotation are characterized by $D_\mr{f}\approx 1.5$, while
these formed without rotation have $D_\mr{f} \approx 2.0$) seem to
behave in a similar way indicating that the main parameter is the
filling factor of the projectile, not the precise structure. We note
that the final increase of the relative toothing radius for CCA
aggregates in \fg{tooth}b and flattening in \fg{tooth}a are probably
caused by chance effects due to the random aggregate shape and the
limited size of the final aggregates (see section~\ref{sec:ch2-rot}).

%
%
% DISCUSSION: ROTATION EFFECT
%
%
\subsection{\label{sec:ch2-rot}The influence of rotation}

We have shown in \se{ch2-methods} that, as long as rotation and linear
motion are associated with similar energies, the effect of rotation on
hierarchical growth should be small.  However, the formation of small
CCA projectiles is in fact influenced significantly by rotation, and
we expect that hierarchical growth will reflect the increased
projectile porosity.

Even though the size of aggregates produced in simulations treating
aggregate rotation is limited, we show the effect of including
rotation with an example calculation.  The overall filling factor of
aggregates formed with rotation is shown, along with the filling
factors discussed earlier, in \fg{ff-vs-np}.  Even for the smallest
projectiles (monomers and dimers are structurally identical to the
non-rotation case), there is a weak decrease in filling factor. This
is the direct, limited influence of projectile rotation during the
collision.  The shift in filling factor seems to be approximately
constant for the small projectiles.  However, starting with 8-mers,
the effect is clearly visible, both in the filling factor of the
projectile, and consequentially in the filling factor of the produced
large aggregate.  The difference of the filling factor between growth
using monomers and using 256-mers reaches two orders of magnitude.
Going back to figure \ref{fig:tooth}, it is interesting to see that
the ratio of final to projectile filling factor follows the same trend
already established for the non-rotation simulations. This re-enforces
our notion that for hierarchical growth, rotation can be neglected
when building up the large aggregate from smaller ones, but it cannot
be ignored when constructing the projectiles.

\Fg{CCAdensity-shells-r} shows the density structure of several
aggregates made of different masses of CCA projectiles formed in the
presence of rotation.  Particles made of largest projectiles contain
only a few hundred bullets and thus have not yet fully converged to
the final structure.  Even close to the core the density will still
increase somewhat.
 \figsmall{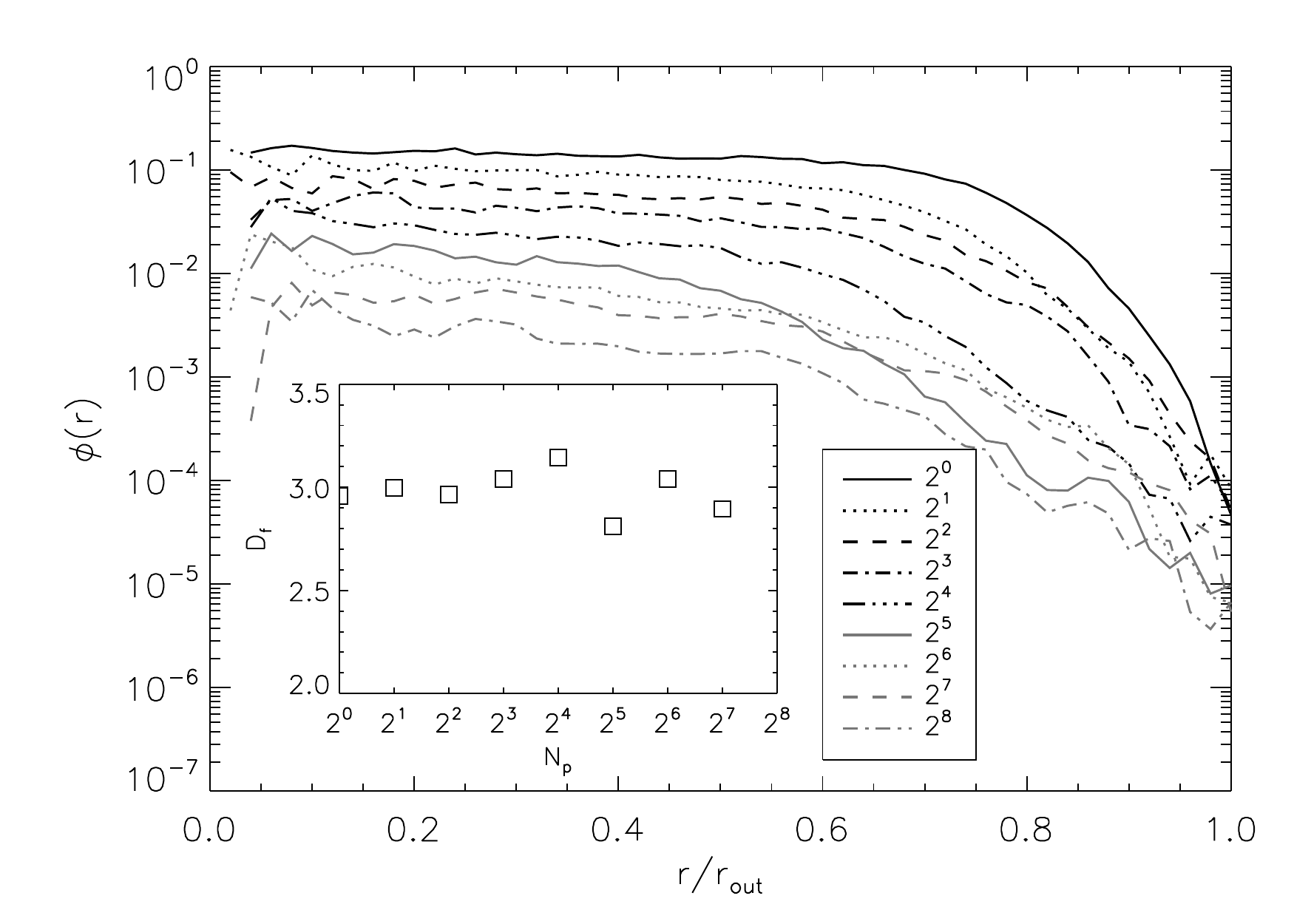}{Density structure of CCA aggregates
   formed with rotation. Different lines corresponds to aggregates formed out of
   projectiles of different masses. From top to bottom projectile masses are:
   $N_\mr{p}=2^0,2^1,2^2,\ldots,2^8$. The inset shows the fractal dimension
   $D_\mr{f}$ determined for these aggregates as a function of the mass of a
   projectile.}{CCAdensity-shells-r}
 Although the density of some aggregates is not homogeneous throughout the inner
 regions, the fractal dimension determined for all aggregates indicates
 non-fractal structure, as presented in the inset of\,
 \fg{CCAdensity-shells-r}. The fractal dimension is scattered around
 $D_\mr{f}=3.0$ and does not fall below $D_\mr{f}=2.7$.
%
%
% RELEVANCE
%
%
\subsection{\label{sec:ch2-relevance}Relevance of the hierarchical growth.}
Hierarchical growth produces aggregates of very porous and non-fractal
structure.  This affects both the collision cross-section and the
strength of particles as the packing density determines the ability to
restructure or erode an aggregate \citep{2008aPaszun}. Since the
influence of the outer fluffy layers of agglomerates on the average
filling factor depends on mass (cf. \se{ch2-surf-effect}), it is
important to understand conditions under which hierarchical growth
takes place. For this purpose we employ a simple model of the
minimum-mass solar nebula \citep{1981PThPS..70...35H} to obtain
quantities that characterize the gas component (\Tb{mdot-model}; after
\citet{2002ApJ...581.1344T} and \citet{2008ApJ...679.1588O}).
\begin{table*}[!ht]
\caption{Gas parameters corresponding to a minimum-mass solar nebula model with
  total mass of $2.5\ep{-2} M_\odot$ within 100AU. The radial distribution of
  the surface density and the temperature have a power-law form with slope of
  $-1$ and $-0.5$, respectively.}  \label{tab:mdot-model}
\begin{tabular}{l c c c}
\hline \hline
parameter & symbol & adopted value & units \\
\hline
Gas density & $\rho_\mr{g}$ & 28./0.16 & $10^{-11}$ g cm$^{-3}$ \\
Sound speed & $c_\mr{s}$ & 10./5.6 & $10^4$ cm s$^{-1}$ \\
Mean free path (gas) & $\lambda$ & 6.9/1230. & cm \\
Temperature & $T$ & 280./89. & K \\
Large eddy overturn time & $\tau_\mr{L}=\Omega^{-1}$ & 0.16/5.0 & yr \\
Turbulence strength parameter & $\alpha$ & $10^{-2}$ & \\
Smallest eddy turnover time & $\tau_\mr{s}$ & 1.3/131 & $10^3$ s \\
\hline
\end{tabular}
\end{table*}
We assume a turbulent state of the gas with the strength parameter
$\alpha=10^{-2}$. The turbulent viscosity is given by
\citep{1973A&A....24..337S}
\begin{equation}
\nu_\mr{T}=\alpha c_\mr{s} H_\mr{g} = \alpha c_\mr{s}^2 \Omega^{-1},
\label{eq:nu-T}
\end{equation}
with $H_\mr{g}$ being the scale height of the gas disk and $\Omega$ being the
local Kepler frequency. The overturn time of the smallest eddy is
\begin{equation}
\tau_\mr{s} = \mr{Re}^{-1/2}\tau_\mr{L},
\label{eq:tau-s}
\end{equation}
with the Reynolds number $\mr{Re}=\nu_\mr{T}/\nu_\mr{m}$, where $\nu_\mr{m} =
c_\mr{s} \lambda /2$ is the molecular viscosity \citep{1993Icar..106..102C}.
Here we consider only two main sources of relative velocities between dust
particles:
\begin{enumerate}
\item{\emph{Thermal (Brownian) motions.}} Dust particles collide with
  gas molecules resulting in transfer of the momentum. These
  collisions give a push from random directions causing Brownian
  motion of dust particles.  Relative velocities between two aggregates
  of mass $m_1$ and $m_2$ are then given by
\begin{equation}
\Delta v_\mr{BM} = \sqrt{\frac{8 \pi k_\mr{B} T (m_1+m_2)}{\pi m_1 m_2}},
\label{eq:brown}
\end{equation}
where $k_\mr{B}$ is Boltzmann's constant. These relative velocities
are very low ($\sim$mm/s for micron-sized particles) and decrease
further for larger sizes.
\item{\emph{Turbulence.}} Dust aggregates respond to the gas on a timescale
  given by the stopping time $\tau_\mr{f}$ (see \eq{tau-f}).  The mean turbulent
  gas velocity fluctuations $v_\mr{g} = (3/2)^{1/2}\ \alpha^{1/2}\ c_\mr{s}$
  \citep{2003Icar..164..127C,2008ApJ...679.1588O} are not immediately mirrored
  by dust particles, resulting in relative motions between dust particles as
  particles of different stopping time have different velocities. These relative
  velocities are given in a simple form \citep{1980A&A....85..316V} as $\Delta
  v_{12}^2=\Delta v_\mr{II}^2 + \Delta v_\mr{I}^2$ where subscripts II and I
  denote contributions due to fast (class II) and slow (class I) eddies.  The
  two terms are given as a function of the Stokes number of the particles by
  \citep[for a complete derivation see][]{2007A&A...466..413O}
\begin{subequations}
\begin{equation}
\Delta v_\mr{I}^2 = v_\mr{g}^2 \frac{\St_1-\St_2}{\St_1+\St_2} \Biggl(
\frac{\St_1^2}{\St_{12}^*+\St_1} - \frac{\St_1^2}{1+\St_1}-(1\leftrightarrow 2)
\Biggr) \label{eq:v-I}
\end{equation}
\begin{equation}
\Delta v_\mr{II}^2 = v_\mr{g}^2 \Biggl( (\St_{12}^*-\mr{Re}^{-1/2})+
\frac{\St_1^2}{\St_1+\St_{12}^*} - \frac{\St_1^2}{\St_1+\mr{Re}^{-1/2}} +
(1\leftrightarrow 2) \Biggr). \label{eq:v-II}
\end{equation}
\label{eq:v-III}
\end{subequations}

\noindent Here the term $\St_{12}^*$ is obtained by solving eq.21d from
\citet{2007A&A...466..413O} and is given by $\St_{12}^* \approx 1.6
\frac{\mr{max}(\tau_1,\tau_2)}{\tau_\mr{L}}$ for small particles ($\tau_\mr{L}
\gg \tau_\mr{f}$), and $\St_{12}^* \approx \frac{\mr{max}(\tau_1,
  \tau_2)}{\tau_\mr{L}}$ for $\tau_\mr{f} \approx \tau_\mr{L}$.
\end{enumerate}

We use these velocities to determine under what circumstances
hierarchical growth is relevant. First we discuss which particles of a
given size distribution contribute most to the growth of a larger
aggregate.  Then we also consider typical impact energies to infer if
and for what particles our idealized hit-and-stick assumptions are
applicable.

\subsubsection{Small particle contributions to growth}

In \se{ch2-introduction} we have already motivated that hierarchical
growth needs to be considered in a situation where on-going
fragmentation leads to a replenishment of small particles.  To
investigate the mass contribution of different size aggregates to the
growth, we will assume a \emph{flat} mass spectrum of dust particles,
i.e.  a distribution that is given by $f(m)m^2\mr{d}\log m =
\mr{const}$.  In this case, the mass in a logarithmic interval between
$\log m$ and $\log m+ \mr{d}\log m$ is constant.  This assumption is a
reasonable approximation to the results of numerical simulations
including fragmentation
\citep[\eg][]{2008A&A...480..859B,2008OrmelMolcloud}. The mass
accumulation rate of a target aggregate is then given by
\begin{equation}
\frac{\mr{d}m_\mr{t}}{\mr{d}t} = \int_{m_1}^{m_2} f_\mr{p}(m)\ m_\mr{p}^2\
\sigma_\mr{coll}\ \Delta v\ \mr{d}m , \label{eq:mdot}
\end{equation}
where $f_\mr{p}(m)$ is density of projectiles of mass $N_\mr{p}$ and
$\sigma_\mr{coll} = \pi (r_\mr{p}+r_\mr{t})^2$ is the cross-section for
collision between target aggregate and a projectile. \Fg{mdot}
\figsmall{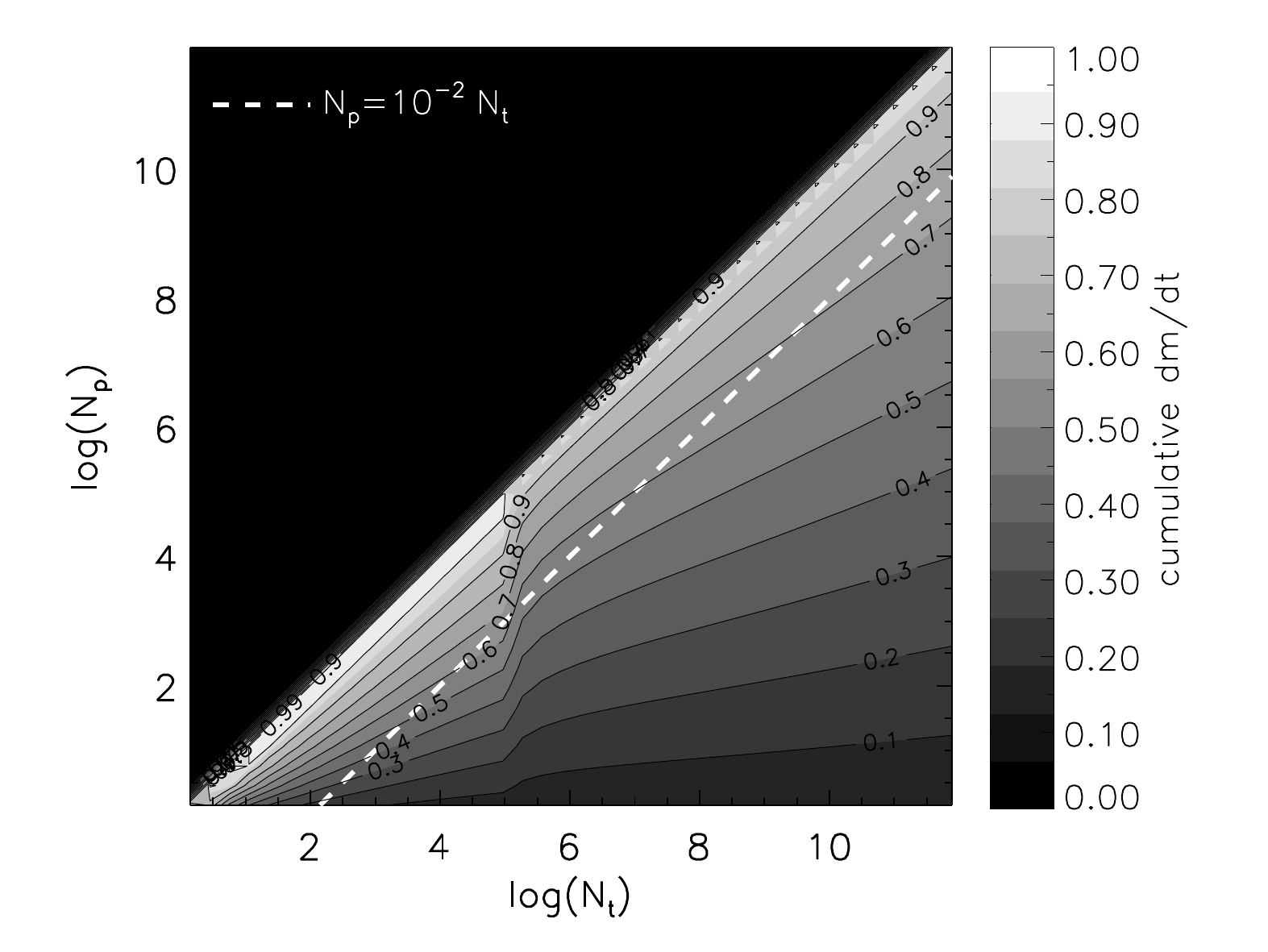}{Cumulative mass accumulation rate for PCA
  aggregates. The mass gain is normalized such that the total accumulation is
  unity for all projectiles smaller or equal to the target mass.  The dashed
  line indicates the projectile size that has two orders of magnitude smaller
  mass than the target ($N_\mr{p} = 10^{-2} N_\mr{t}$). The change in the mass
  accumulation rate at about $N_\mr{t} \approx 10^5$ indicates a change in
  relative velocities due to decoupling of aggregates from the smallest
  eddies.}{mdot}
shows the \emph{cumulative} mass accumulation rate
\textit{\.m$_\mr{t}$} for every target aggregate. For each target
($N_\mr{t}$) and projectile ($N_\mr{p}$) we show the mass contribution
from all projectiles smaller or equal to $N_\mr{p}$.

%% The moment of the transition at about $N=10^5$ indicates decoupling of
%% particles from the smallest eddies. This obviously depends on the filling
%% factor, as the projected surface (and thus the stopping time) is tightly
%% related to how compact the particle is \begin{equation}
%% \sigma=\biggl(\frac{N}{\ff} \biggr)^{2/3} r_0^2.
%% \label{eq:sigma-phi-rel} \end{equation} The more fluffy the particle is, the
%% more it is coupled to the gas, as it has larger cross-section for interaction
%% with the gas molecules.

Given these assumptions, \Fg{mdot} clearly shows that for all but the
smallest aggregates, the growth is dominated by projectiles much
smaller that the target.  The dashed line indicates a projectile mass
two orders of magnitude smaller than the corresponding target
particle.  Therefore aggregates made of more than $\sim 10^5$ monomers
collect over 60\% of the total coagulated material from particles of
masses smaller than 0.01 of their own. This indicates a great
importance of hierarchical growth, as the coagulation is dominated by
collisions with significantly smaller particles.

Prediction of the aggregate structure that results from the
hierarchical growth, requires a coagulation model that follows the
evolution of both the mass and the filling factor of dust
aggregates. This issue is going to be the subject of a follow-up
study.

\subsubsection{Impact energy and the hit-and-stick approximation}

As the hierarchical growth has a great importance in the coagulation
of dust aggregates, we now need to focus on the energy available
during a collision and verify under what conditions this energy is
consistent with the assumptions made in this paper.  We have shown in
\se{ch2-pca} and \se{ch2-cca} that the structure of aggregates in the
hit-and-stick regime is determined almost entirely by the filling
factor of projectiles.  This opens up a very simple way to compute
filling factors of aggregates in this regime.  Higher collision
energies will result in restructuring and thus affect the final
porosity of aggregates (making them more compact).  The restructuring
threshold depends on both material properties and monomer size
(cf. \eq{Erol}).  We keep these quantities as free parameters in the
following derivation.  The surface energy $\gamma$ for two species we
adopt is $\gamma_\mr{sil}=25$ erg cm$^{-2}$ for silicate monomers and
$\gamma_\mr{ice} = 370$ erg cm$^{-2}$ for ice coated silicate
grains. In the latter case the mass of aggregates is set by the bulk
density of silicates ($\rho_0 = 2.65$ g cm$^{-3}$), while the
restructuring is determined by the ice mantle.  We also consider three
different monomer sizes: $r_0 = 300\,\AA$, $r_0 = 0.1\,\mu$m, and $r_0
= 0.5\,\mu$m.

Numerical simulations by \citet{1997ApJ...480..647D} have shown that
the first visible restructuring occurs above about five times the
rolling energy (\eq{Erol}).  Later \citet{2000Icar..143..138B}
confirmed this finding in laboratory experiments.  Therefore we take
this energy threshold as the upper limit of the hit-and-stick regime.
As for the relative velocities we continue with the minimum-mass solar
nebula model adopted in \se{ch2-relevance}.

We calculate the collision energies between different particles
relative to the restructuring energy threshold: $E/(5\ E_\mr{roll})$.
Note that the rolling energy $E_\mr{roll}$ scales for different
materials (ice and silicates) with the surface energy $\gamma$ only,
as the critical displacement $\xi_\mr{crit}$ is assumed to be
independent of material.  \Fg{rel-vel} shows the projectile mass for
which the impact energy equals the threshold energy for restructuring.
The different line styles indicate aggregates
\figsmall{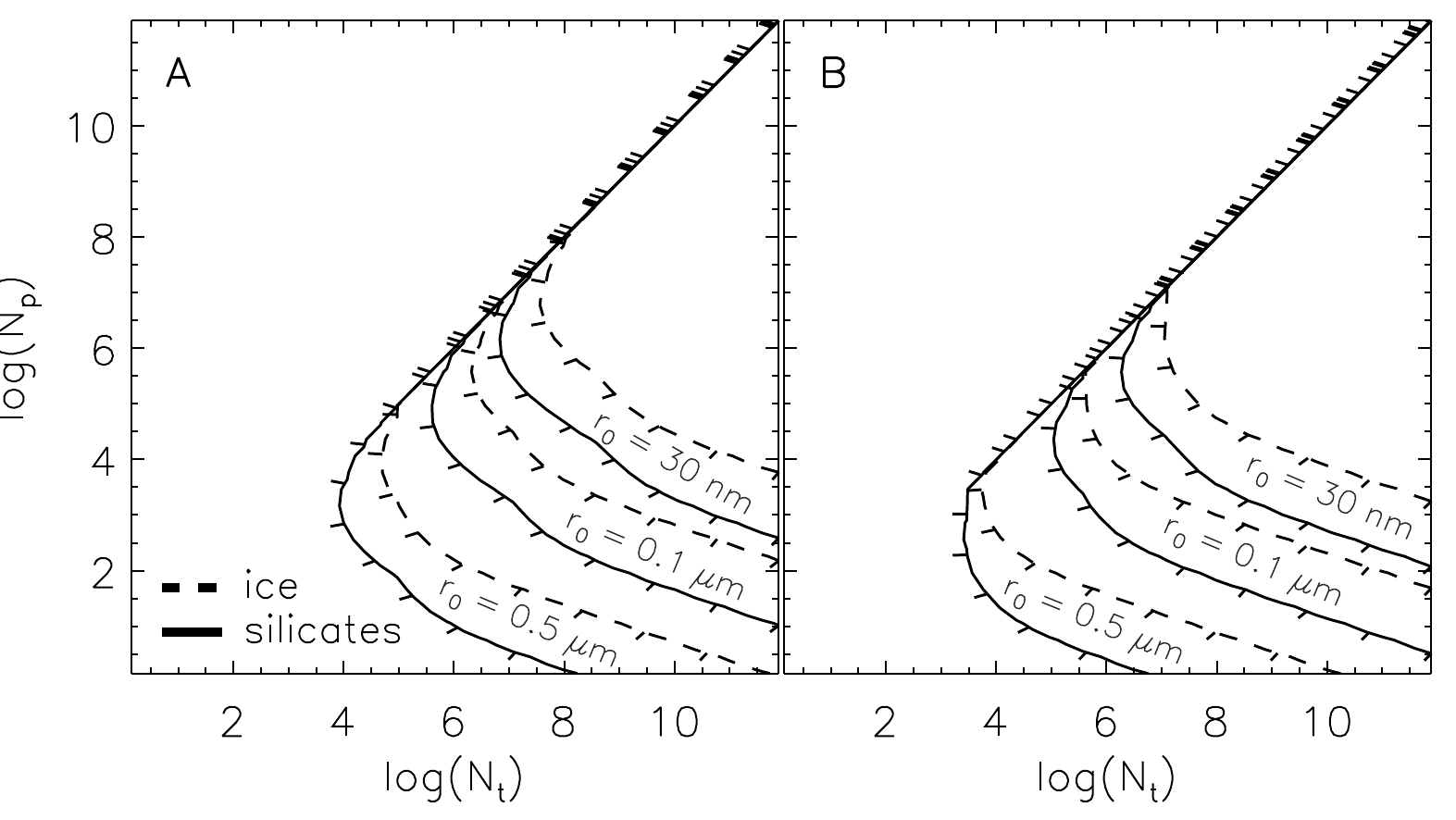}{The restructuring energy threshold
  ($E=5E_\mr{roll}$) for small monomers of $r_0=300\ \AA$, $r_0=0.1$, and
  $r_0=0.5$ $\mu$m. The solid line correspond to silicate grains, while the
  dashed line corresponds to ice coated silicate grains. Short perpendicular
  ticks indicate a decrease direction of the collision energy. The two panels
  correspond to different distances from the central star in the minimum-mass
  solar nebula: A - 1AU and B - 10AU.}{rel-vel}
made of ice-coated monomers (dashed lines) and silicates grains (solid
lines).  Lines corresponding to different monomer sizes are labeled.
Collisions occur between target aggregates (of mass $N_\mr{t}$) and
projectiles ($N_\mr{p}$).  For simplicity we assume a PCA structure of
all aggregates, i.e. $\ff=0.15$, which is really the worst case
scenario. If more fluffy particles form due to hierarchical growth,
the porosity will increase, changing the projected surface area of
agglomerates.  This affects both their collision cross-section and
their stopping time.  As a result, relative velocities will decrease,
allowing hit-and-stick growth to proceed to even larger
masses. \Fg{rel-vel}a shows the results at the distance of 1 AU from
the central star.  Aggregates made of about micron-sized silicate
monomers ($r_0 = 0.5\, \mu$m) can grow to about $N_\mr{t}=10^4$
grains, before any restructuring can occur.  Smaller aggregates, of
the order of $N_\mr{p} \sim 10$ monomers, have impact energies above
the threshold value when they collide with very large targets made of
over $N_\mr{t} \sim 10^6$ grains.  A significantly higher threshold
velocity for smaller monomers allows for the hit-and-stick growth up
to greatly larger aggregate sizes (in terms of the number of
monomers).  No restructuring occurs, regardless of the projectile
mass, for target aggregates made of, respectively, up to over 10$^6$
and up to over 10$^7$ monomers for grain size of 0.1 $\mu$m and 300
$\AA$.  This effect is caused by both higher adhesion force per unit
mass (increasing the rigidity of aggregates), and much shorter
stopping times of aggregates made of smaller monomers (leading to
smaller relative velocities). The effect of the difference in the
rolling energy can be directly observed by comparing results for
silicate grains with and without ice mantle. The higher surface energy
of ice results in a threshold energy higher by about an order of
magnitude (this is caused by the ratio of the surface energy of ice
over that of silicates of $\gamma_\mr{ice}/\gamma_\mr{sil} = 14.8$)
for the hit-and-stick growth.

\Fg{rel-vel}b shows the threshold energy for aggregates at 10 AU. In
this case, lower densities (due to the negative radial density
profile) result in longer stopping times of aggregates and thus higher
relative velocities. The onset of restructuring at 10 AU, however, is
only slightly different. The maximum number of monomers that can be
reached without the onset of restructuring is only slightly smaller.

These results show that dust aggregates can grow several orders of
magnitude in mass before any restructuring can occur. In the case of a
weaker turbulence the relative velocities decrease. This results in
the growth unaffected by restructuring to even larger sizes. For the
parameter $\alpha = 10^{-4}$ particles will grow to masses about two
orders of magnitude larger. The coagulation process in these
conditions produces particles of very low filling factor, depending on
the mechanism responsible for the growth of small projectiles.

Note that small aggregates can also collide with much larger particles
(due to a flat mass spectrum) and cause restructuring. In such a case,
however, these small aggregates are incorporated into bigger bodies
(they are removed) and restructuring occurs only in this large
particle. Moreover, very big particles (about 100$\mu$m in size and
larger) can bounce \citep{2007arXiv0711.2148L}, which affects the
structure of both impactors if the impact energy is sufficiently
high. This, however, requires growth to very big sizes, where
restructuring is already involved in shaping the particle
structure. Therefore, the hierarchical growth in this size regime must
be considered together with a structural evolution due to collisional
compaction \citep[\eg][]{2008aPaszun}.

\section{\label{sec:ch2-conclusions}Conclusions}

We have studied the process of hierarchical growth of dust aggregates,
in which a large target aggregate is built up slowly by adding much
smaller projectiles of constant size.  We have shown that this process
leads to aggregates that consist of a core and a surface transition
region.

The core is non-fractal in nature, so it has a constant density.  The
density of this core results from the fact that it is built up in a
PCA-like process from the projectiles.  If the projectiles are
particles that are relatively compact and spherical (i.e. PCA
aggregates themselves), a good approximation for the final core
density is the product of the volume filling factor of the projectiles
$\ff_\mr{p}$ times the PCA filling factor 0.15.  For projectile
aggregates that result from a CCA-like growth mechanism, the typical
distance between projectile aggregates in the growing target is
smaller than the circumscribing radius of the projectile, because the
porous outer layers interpenetrate before the first physical contact.
However, also in this case, we find that the final filling factor of
the core is largely given by the filling factor of the projectiles and
can be approximated very well by a simple relation (see \eq{phi-fit}).

We also show that in a coagulation environment where relative
velocities are driven by turbulence, a logarithmically flat mass
distribution (equal mass per mass decade) immediately leads to a
situation where small particles and aggregates dominate the growth of
large ones. Therefore, in such environments, hierarchical growth
should be seen as the norm. Consequently, we predict that the
aggregates in such environments are not fractals with extremely low
densities as they would result from extrapolation of fractal laws to
large sizes.  The compactification of aggregates does not only result
from collisions with enough energy to restructure aggregates - it
starts already earlier by filling voids in particles with smaller
projectiles that contribute to the growth.

\section*{Acknowledgments}
We thank Prof. J\"urgen Blum for useful comments, Evghenii Gaburov for
constructive discussions on the implementation of the nearest neighbor
search algorithm, and Chris Ormel for input on the velocity field of
particles in a disk model.  We also thank SARA supercomputer center
for providing us with access to computer cluster LISA, to run very
time consuming simulations.  Travel support of the Leids
Kerkhoven-Bosscha Fonds has been important for this research, which
was financed by the Nederlandse Organisatie voor Wetenschapelijk
Onderzoek, Grant 614.000.309.

\bibliographystyle{aa} 

%\bibliography{../citation}
\end{document}